# Magnetotransport in Weyl semimetal WTe$_2$ single crystal

A. N. Perevalova[1]✉, B. M. Fominykh[1,2], V. V. Chistyakov[1], S. V. Naumov[1], V. N. Neverov[1], V. V. Marchenkov[1,2]

[1]M.N. Mikheev Institute of Metal Physics, UB RAS, Ekaterinburg, Russia
[2]Ural Federal University, Ekaterinburg, Russia
✉domozhirova@imp.uran.ru

**Abstract.** A WTe$_2$ single crystal was grown by the chemical vapor transport method, and its electrical resistivity and galvanomagnetic properties were investigated. Single-band and two-band models were used to estimate the concentration and mobility of charge carriers in WTe$_2$ at temperatures from 4.2 to 150 K.

**Funding:** The results of studies of magnetoresistivity were obtained within the state assignment of the Ministry of Science and Higher Education of the Russian Federation (theme "Spin," No. 122021000036-3), supported in part by the scholarship of the President of the Russian Federation to young scientists and graduate students (A.N.P., SP-2705.2022.1). One of us (V.V.M.) appreciates the support of the Ural Federal University (Priority-2030 Program).

## Introduction

In recent years, WTe$_2$ has attracted great interest due to its extremely large magnetoresistance, which varies with the field according to a law close to quadratic without saturation up to 60 T [1]. It was assumed that the mechanism leading to the extremely large magnetoresistance in WTe$_2$ is electron-hole compensation [1]. Theoretical calculations of the electronic structure of WTe$_2$, as well as experimental studies using the angle-resolved photoemission spectroscopy (ARPES) and Shubnikov-de Haas oscillation measurements, revealed that at low temperatures the Fermi surface of this compound has a complex structure and consists of several pairs of electron-like and hole-like pockets, while the total volume of electron pockets is equal to that of hole pockets [1-4].

Another interesting feature of WTe$_2$ is that it was predicted as a candidate type-II Weyl semimetal [4], which was confirmed experimentally using ARPES and scanning tunneling microscopy [5, 6]. In such materials, the valence and conduction bands with a linear dispersion touch at points near the Fermi level, forming the so-called Weyl nodes. Instead of a point-like Fermi surface characteristic of a type-I Weyl cone, the type-II Weyl node is a touching point of the electron and hole pockets. Quasiparticles near such nodes behave similarly to massless Weyl fermions in high-energy physics. Weyl points always occur in pairs of opposite chirality and are topologically protected, resulting in unique surface states called Fermi arcs and unusual transport properties [7, 8]. The low effective mass of current carriers in Weyl semimetals leads to their high mobility ~$10^3$-$10^6$ cm$^2$/(V·s), which opens up prospects for the use of such materials for the development of ultrafast electronic devices.

The concentration and mobility of current carriers can be estimated based on data on the Hall effect and magnetoresistivity. On the one hand, such estimates are carried out within the framework of a single-band model, in particular, in [9, 10], where the transport properties of Weyl and Dirac semimetals were studied. On the other hand, to analyze the galvanomagnetic properties of such materials, the two-band model is often used, which takes into account the contributions of both electron and hole carriers [11, 12]. In [13], we compared the results of using these models to estimate the concentration and mobility of current carriers in $WTe_2$ at $T = 12$ K and obtained good agreement. However, it is of interest to carry out such an analysis over a wider temperature range. The aim of this work is to compare the results of using the single-band and two-band models for the analysis of the galvanomagnetic properties of $WTe_2$ in the temperature range from 4.2 K to 150 K.

## Materials and Methods

$WTe_2$ single crystals were grown by the chemical vapor transport method, which was described in detail in [13]. The crystal structure of the synthesized crystals was studied by X-ray diffraction using $CrK\alpha$ radiation. It was found that the single crystals under study have an orthorhombic structure (space group $Pmn2_1$) with lattice parameters $a = 3.435(8)$ Å, $b = 6.312(7)$ Å, $c = 14.070(4)$ Å. Using X-ray energy-dispersive microanalysis on a Quanta 200 Pegasus scanning electron microscope with an EDAX attachment, it was established that the chemical composition of the grown crystals corresponds to stoichiometric $WTe_2$.

To measure the transport characteristics, a sample with dimensions of ~$4 \times 1 \times 0.4$ mm$^3$ was chosen. The resistivity and Hall resistivity were measured by the four-contact method in the temperature range from 4.2 K to 290 K and in magnetic fields up to 9 T on an Oxford Instruments setup for studying galvanomagnetic phenomena in strong magnetic fields and under low temperatures. During measurements, the electric current flowed in the $ab$ plane, and the magnetic field was directed along $c$ axis. To estimate the quality of the sample, the residual resistivity ratio (RRR) was determined, which is the ratio of resistivities at room temperature and at liquid helium temperature and depends on the number of defects and impurities. In this work, the RRR value for the single crystal under study is $\rho_{300\,K}/\rho_{4.2\,K} \approx 50$, which is comparable with the RRR in [14], but at the same time, it is less than in [1, 11].

## Results and Discussion

Fig. 1(a) shows the temperature dependence of the electrical resistivity $\rho(T)$ of the $WTe_2$ single crystal in a zero magnetic field and in a field of 9 T. It can be seen that in the absence of a magnetic field, the sample exhibits a metallic behavior with an increase in the electrical resistivity from $0.13 \cdot 10^{-4}$ Ohm·cm to $5.7 \cdot 10^{-4}$ Ohm·cm with increasing temperature from 4.2 K to 290 K. While the magnetic field $B = 9$ T leads to the appearance of a minimum in the dependence $\rho(T)$ at $T \approx 60$ K. It is assumed that this minimum can be caused by the transition from effectively high ($\omega_c\tau \gg 1$, where $\omega_c$ is the cyclotron frequency, $\tau$ is the relaxation time) to weak ($\omega_c\tau \ll 1$) magnetic fields [15, 16], which is observed for compensated conductors with a closed Fermi surface [17]. The term "compensated" implies that the concentrations of electrons and holes are equal in such materials. To estimate the concentration $n$ as well as mobility $\mu$ of charge carriers, a single-band model is often used for the analysis of data on the Hall effect, where $n = \frac{1}{R_H \cdot e}$, $\mu = \frac{R_H}{\rho}$ (here $R_H$ is the Hall coefficient; $e$ is the electron charge; $\rho$ is the electrical resistivity in a zero magnetic field). Fig. 1(b) shows the temperature dependence of the Hall coefficient $R_H$ of WTe$_2$ in a field $B = 9$ T. It can be seen that $R_H$ has a negative sign, that is, electrons are the main charge carriers in the single crystal under study. Inserts in Fig. 1(b) show the temperature dependences of the concentration $n$ and the mobility $\mu$ of current carriers in WTe$_2$, obtained using the single-band model. At $T = 4.2$ K, $n$ and $\mu$ are $5 \cdot 10^{19}$ cm$^{-3}$ and $7 \cdot 10^3$ cm$^2$/(V·s), respectively, where $n$ increases and $\mu$ decreases with increasing temperature.

At the same time, since it is known [1-4] that WTe$_2$ is a compensated semimetal, it is of interest to estimate the concentrations and mobilities of separately electron and hole current carriers. In this case, a two-band model is used, where the resistivity $\rho_{xx}$ and the Hall resistivity $\rho_{xy}$ can be represented as [11]:

$$\rho_{xx} = \frac{1}{e} \frac{(n_h\mu_h + n_e\mu_e) + (n_h\mu_e + n_e\mu_h)\mu_h\mu_e B^2}{(n_h\mu_h + n_e\mu_e)^2 + (n_h - n_e)^2 \mu_h^2 \mu_e^2 B^2}, \quad (1)$$

$$\rho_{xy} = \frac{B}{e} \frac{(n_h\mu_h^2 - n_e\mu_e^2) + (n_h - n_e)\mu_h^2\mu_e^2 B^2}{(n_h\mu_h + n_e\mu_e)^2 + (n_h - n_e)^2 \mu_h^2 \mu_e^2 B^2}. \quad (2)$$

Here $n_e$ ($n_h$) and $\mu_e$ ($\mu_h$) are the concentration and mobility of electrons (holes), respectively. From the simultaneous fitting of the experimental curves $\rho_{xx}(B)$ and $\rho_{xy}(B)$ in the framework of the two-band model, one can obtain the mobilities and concentrations of electron and hole current carriers. Fig. 2(a) shows the $\rho_{xx}(B)$ and $\rho_{xy}(B)$ dependences for WTe$_2$. Open symbols correspond to experimental data, and solid lines correspond to curves obtained within the framework of the two-band model using a computer program [18]. It can be seen that the fitting curves describe the experimental results well. The error in determining the fitting parameters does not exceed 3%. Fig. 2(b) shows the temperature dependences of the concentrations and mobilities of electrons and holes obtained using the two-band model. At $T = 4.2$ K, concentrations of electrons and holes are $4 \cdot 10^{19}$ cm$^{-3}$ and $3.4 \cdot 10^{19}$ cm$^{-3}$, respectively, which is close to the value of $n$, obtained using the single-band model. At temperatures above 50 K, the electron concentration increases, while the hole concentration decreases. Whereas within the framework of the single-band model, an increase in the concentration of the main carriers is observed. At $T = 4.2$ K, the mobilities of electrons and holes are $6.3 \cdot 10^3$ cm$^2$/(V·s) and $3.7 \cdot 10^3$ cm$^2$/(V·s), respectively, and decrease with increasing temperature, which agrees with the results obtained using the single-band model. Note that the concentration and mobility of electrons mainly exceed that of holes, which means that electrons are the majority carriers, which is also consistent with the results of the single-band model. The inset in Fig. 2(b) shows that the ratio $n_h/n_e$ is close to 1 at temperatures below 50 K, which indicates a state close to electron-hole compensation in WTe$_2$. It was also shown in [11] by analyzing data on the Hall effect and magnetoresistivity in WTe$_2$ that the concentrations of electrons and holes are comparable at $T < 50$ K, which may be due to a change of the electron structure of this compound at low temperatures.

According to the Eq. (1), the magnetoresistivity (MR) $\frac{\Delta \rho}{\rho}$ can be expressed by the formula:

$$\frac{\Delta \rho}{\rho} = \frac{\rho_{xx} - \rho}{\rho} = \frac{(n_h \mu_h + n_e \mu_e)^2 + (n_h \mu_h + n_e \mu_e)(n_h \mu_e + n_e \mu_h) \mu_h \mu_e B^2}{(n_h \mu_h + n_e \mu_e)^2 + (n_h - n_e)^2 \mu_h^2 \mu_e^2 B^2} - 1. \quad (3)$$

Thus, if the compensation condition is satisfied, $n_e = n_h$, the MR can be represented as $\frac{\Delta \rho}{\rho} = \mu_e \mu_h B^2$, that is, it changes with the field according to a quadratic law. Fig. 3 shows a plot of the MR versus magnetic field $B$ for WTe$_2$ at $T = 4.2$ K. The MR reaches 1750% in a field of 9 T, which is less than the MR in [1, 11], which is apparently due to the lower value of the RRR for our crystal. Approximation of this dependence by a power function (red solid line in Fig. 3) revealed that $\frac{\Delta \rho}{\rho} \sim B^{1.87}$, i.e. at $T = 4.2$ K, the MR changes with the field according to a nearly quadratic law. The triangular symbols in Fig. 3 show the MR calculated by the Eq. (3) for the case $n_e = n_h$, where the values of $\mu_e$ and $\mu_h$ were obtained using the two-band model. It can be seen that the MR calculated from the Eq. (3) exceed those obtained from the experimental data at $B \geq 5$ T. Apparently, this is due to non-ideal compensation in the WTe$_2$ single crystal under study.

## Conclusion

The WTe$_2$ single crystal was grown and its galvanomagnetic properties were studied. Hall effect data as well as magnetoresistivity data were analyzed using both single-band and two-band models, followed by extracting the concentration and mobility of current carriers. It can be assumed that the results obtained within the framework of the two-band model are in good agreement with the results of applying the single-band model. Moreover, the two-band model, which provides information on the concentration and mobility of both electron and hole carriers, is preferable for systems containing different groups of carriers.

## Acknowledgments

Structural and transport studies were carried out using equipment of the Collaborative Access Center "Testing Center of Nanotechnology and Advanced Materials" of the Ural Branch of the Russian Academy of Sciences.

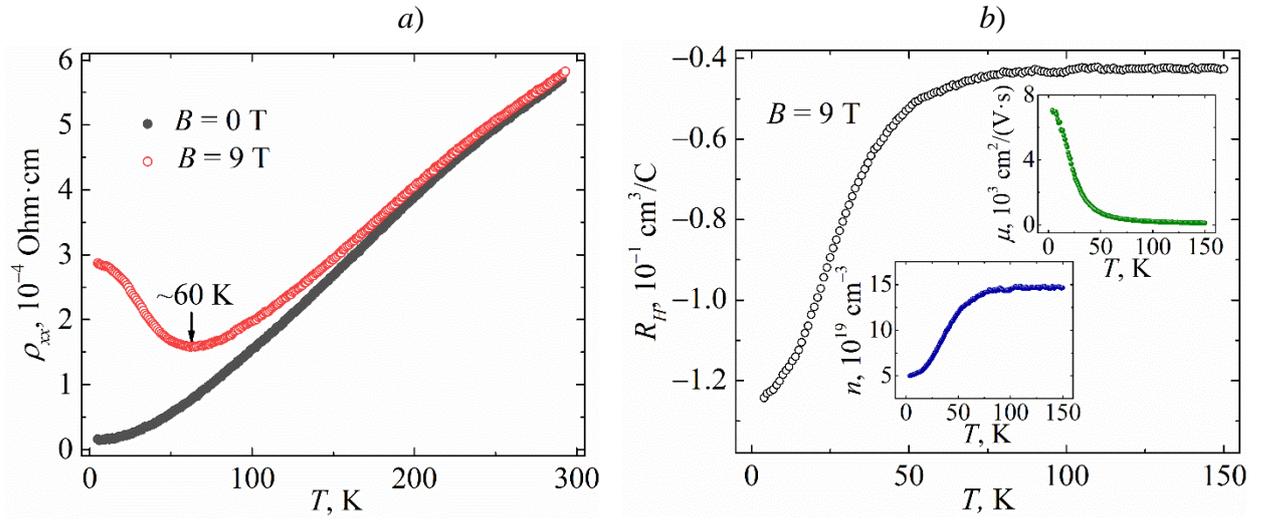

Fig. 1. (a) Temperature dependence of electrical resistivity in a zero magnetic field and in a field of 9 T. (b) Temperature dependence of the Hall coefficient $R_H$ of WTe$_2$ in a field $B = 9$ T. Inserts show the concentration $n$ and mobility $\mu$ of current carriers versus temperature, obtained using a single-band model.

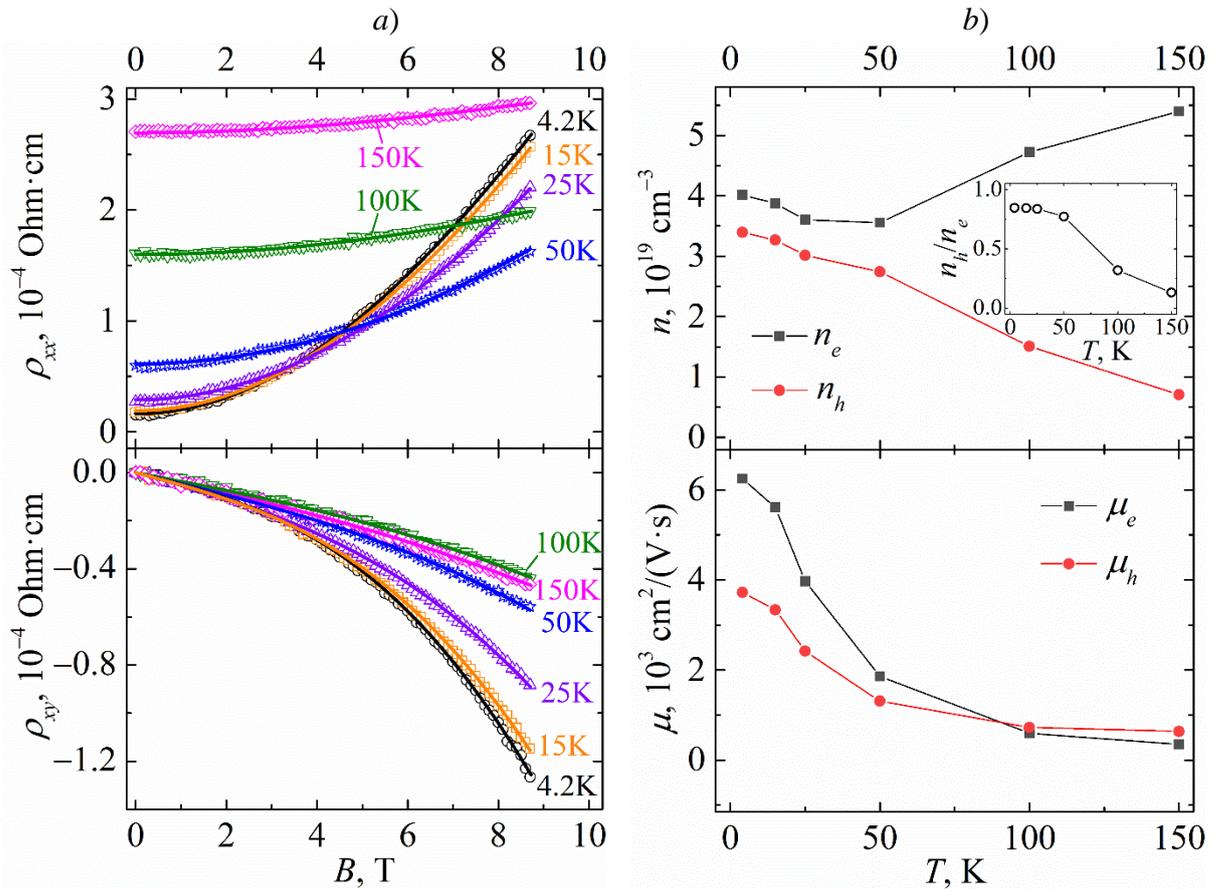

Fig. 2. (a) Field dependences of the resistivity $\rho_{xx}$ and Hall resistivity $\rho_{xy}$ for WTe$_2$ at temperatures from 4.2 K to 150 K: open symbols are experimental data; solid lines are fitting curves obtained using Eq. (1) and (2). (b) Temperature dependences of the concentrations and mobilities of electrons and holes extracted using two-band model. The inset shows the temperature dependence of the ratio $n_h/n_e$.

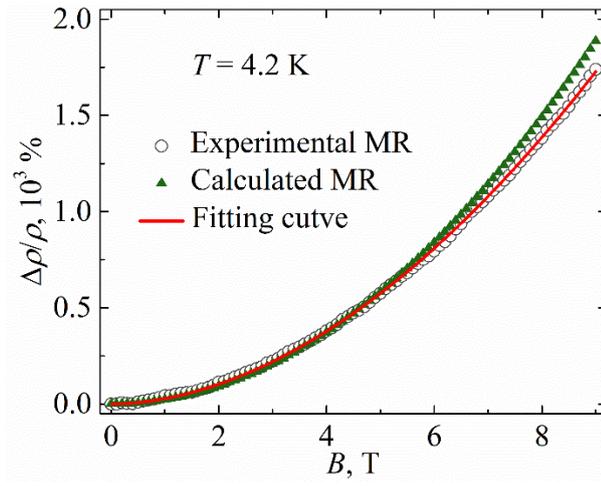

Fig. 3. Field dependence of the magnetoresistivity (MR) of WTe$_2$ at $T = 4.2$ K: open symbols represent the MR calculated from experimental data; the red solid line is the approximating function; triangles are the MR calculated from the compensation condition within the framework of the two-band model.